\documentstyle[prl,aps,floats,epsf,twocolumn]{revtex}
\begin{document}
\flushbottom
\twocolumn[\hsize\textwidth\columnwidth\hsize\csname @twocolumnfalse\endcsname

\title{\bf Stiff monatomic gold wires with a spinning zigzag geometry}

\author{
   Daniel S\'anchez-Portal$^1$,
   Emilio Artacho$^2$,
   Javier Junquera$^2$,
   Pablo Ordej\'on$^3$,
   Alberto Garc\'{\i}a$^4$, and
   Jos\'e M. Soler$^2$
}
\address{
   $^1$Department of Physics and Materials Research Laboratory,
       University of Illinois, Urbana, Illinois 61801 \\
   $^2$Dep.\ de F\'{\i}sica de la Materia Condensada and 
       Inst.\ Nicol\'as Cabrera, C-III,
       Universidad Aut\'onoma, E-28049 Madrid, Spain \\
   $^3$Institut de Ci\`encia de Materials de Barcelona (CSIC)
       Campus de la U.A.B. E-08193 Bellaterra, Barcelona, Spain \\
   $^4$Departamento de  F\'{\i}sica Aplicada II,
       Universidad del Pais Vasco, Apdo.\ 644, E-48080 Bilbao, Spain \\
}

\date{\today}
\maketitle

\begin{abstract}
   Using first principles density functional calculations, 
gold monatomic wires are found to exhibit a zigzag shape which
remains under tension, becoming linear just before breaking. 
   At room temperature they are found to spin, what explains
the extremely long apparent interatomic distances shown by electron
microscopy.
   The zigzag structure is stable if the tension is relieved, the wire
holding its chainlike shape even as a free-standing cluster.
   This unexpected metallic-wire stiffness stems from the 
transverse quantization in the wire, as shown in a simple free 
electron model.
\end{abstract}



]

\par
   The manipulation of matter at the atomic scale~\cite{Eigler} is 
heralding a technological revolution and opening new research avenues.  
   A spectacular achievement is the recent 
fabrication~\cite{Takayanagi,AgraitRuitenbeek} of monatomic 
chains of gold atoms, the ultimate thin wires.  
   Metallic nanowire contacts can be created with the scanning tunneling 
microscope~\cite{Agrait,Pascual,Olesen}, with mechanically controllable
break junctions~\cite{Muller-Krans-Scheer}, or even with simple tabletop
setups~\cite{Costa}.
   The relationships between conduction, geometric, and mechanical 
properties have been studied by simultaneous measurements of 
conductance and applied force~\cite{Rubio}, 
by atomistic~\cite{Landman90,Todorov}, 
continuous~\cite{Torres96,Stafford-Ruitenbeek-Yannouleas-Blom},
or mixed~\cite{Lang,Wang} model simulations, and by first-principles 
calculations~\cite{Landman97,SanchezPortalPRL,Jacobsen99,Torres99}.
   Until very recently, however, only indirect experimental information 
about the structure of the nanocontacts was available.
   This situation changed dramatically after 
Ohnishi {\it et al}~\cite{Takayanagi} directly visualized 
nanometric gold wires by transmission electron microscopy (TEM).
   Surprisingly, in a bridge of four atoms connecting two gold tips,
which was stable for more than two minutes, the atoms were spaced by 
3.5-4.0~\AA.
   Later reports~\cite{TakayanagiAPS} have even increased this distance
up to $\sim$5~\AA, a value much larger than that in Au$_2$ (2.5~\AA) 
and in bulk gold (2.9~\AA). 
   Gold monatomic chains with a length of four or more atoms were
independently associated by Yanson {\it et al}\cite{AgraitRuitenbeek} to 
the last conductance plateau during stretching
(close to one conductance quantum $2e^2/h$).
   The histogram of these plateau lengths showed maxima at regular 
intervals, which might be related to the distances between gold atoms 
in the wire.

\par
   In this work we study the structure and stability of gold
monatomic wires by first-principles density-functional 
calculations~\cite{KohnSham}.
   We use {\sc Siesta}~\cite{Ordejon-SanchezPortalIJQC}, 
a code designed to treat large systems with local basis sets
which has been already used to study gold clusters~\cite{Garzon}.
   Tests were performed for Au$_2$ and bulk gold, using both the
local density approximation (LDA)~\cite{PerdewZunger}
and the generalized gradient approximation (GGA)~\cite{PBE}.
   Core electrons were replaced by scalar-relativistic 
norm-conserving pseudopotentials~\cite{TroullierMartins}.
   Valence electrons were described with a basis set of double-$\zeta$ 
$s, p$ and $d$ numerical pseudo-atomic orbitals. 
   Real- and reciprocal-space integration grids were increased until 
a total-energy convergence better than 2~meV/atom was achieved.
   The results are in very good agreement with previous calculations,
using the same functionals, and with the experimental geometries 
and vibration frequencies~\cite{Aucalculations}.
   The GGA improves the binding and cohesive energies, 
but not the geometries, which are the main focus of this work.
   In the LDA, we obtain, for the gold dimer, a bond length
$l$=2.51~\AA, a vibration frequency $\nu$=190~cm$^{-1}$, 
and a binding energy $D$=3.18~eV.
   For the bulk fcc crystal, the calculated nearest-neighbor distance, 
bulk modulus, and cohesive energy are 
$d$=2.91~\AA, $B$=194~GPa, and $E_c$=4.55~eV respectively.
   In the GGA, the results are 
$l$=2.57~\AA, $\nu$=171~cm$^{-1}$, $D$=2.72~eV, 
$d$=2.98~\AA, $B$=137~GPa, and $E_c$=3.37~eV.
   The experimental values are 
$l$=2.47~\AA, $\nu$=191~cm$^{-1}$, $D$=2.29~eV, 
$d$=2.87~\AA, $B$=172 GPa, and $E_c$=3.78~eV.

\par
   The wire calculations were performed for infinite monatomic chains, 
using periodic boundary conditions, as well as for finite wires
of various lengths, either free-standing or confined between small 
pyramidal tips.
   All the calculations were repeated with the LDA and the GGA, 
and both ferromagnetic and antiferromagnetic solutions were searched.
   In every case, the geometry was relaxed until the maximum 
forces were smaller than 10~meV/\AA\ (16 pN).
   As an additional cross-check, some critical geometries were
recalculated with a different code, using a plane wave basis set.
   The results will be presented in full elsewhere.
   In short, we have found no qualitative differences, and only very 
minor quantitative differences between the finite and infinite wires,
between plane wave and local basis sets, and between LDA and GGA, 
and no magnetic solutions could be stabilized at any wire length.
   We present in what follows the {\sc Siesta} LDA results for the
infinite wires, except where stated.

\begin{figure}[t]
\narrowtext
\centering
\epsfxsize=0.95\linewidth
\epsffile{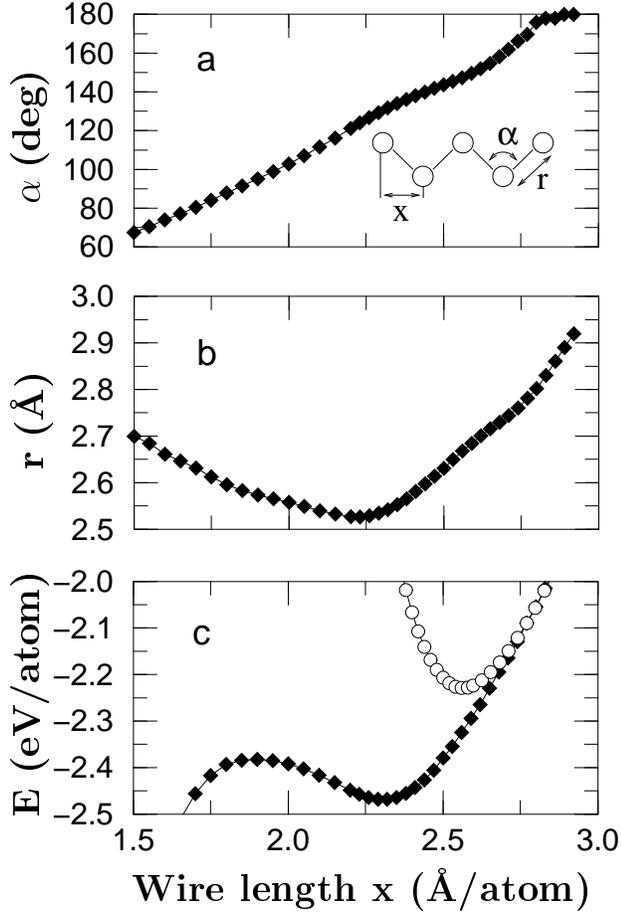}
\vspace{0.5 cm}
\caption[ ] {
First-principles, density-functional results for the
bond angle $\alpha$ (a), and bond length $r$ (b) in a
monatomic gold wire with zigzag geometry, as a function of its
length per atom.
(c) Binding energy $E$ in the zigzag (solid symbols) and
linear wires (open symbols).
}
\label{zigzag}
\end{figure}

\par
   Fig.~\ref{zigzag} shows the wire geometry and the binding energy
as a function of the wire length.
   Except when very stretched, the wire adopts a nonlinear, planar 
zigzag geometry, with two atoms per unit cell.
   Unconstrained relaxations with larger cells did not result in 
longer periods, nor in out-of-plane deformations.
   The energy shows a shallow minimum at a length 
of 2.32~\AA/atom, with a bond angle of 131$^{\rm o}$.
   The stability of this geometry was demonstrated by checking that 
the dynamical matrix, calculated in a cell of 16 atoms, had no 
negative eigenvalues.
   For comparison Fig.~1c shows the energy of
a wire constrained to a linear geometry, which has a minimum
0.24~eV/atom higher, and at a wire length 0.25~\AA\ longer,
than in the zigzag geometry.   
   This difference in wire length is almost entirely due to the change 
in bond angle, since the bond distances differ by only 0.02~\AA\
between the two minima.
   The bond angle increases with stretching, but the wire becomes
linear only shortly before breaking.

\begin{figure}[t!]
\narrowtext
\centering
\epsfxsize=0.95\linewidth
\epsffile{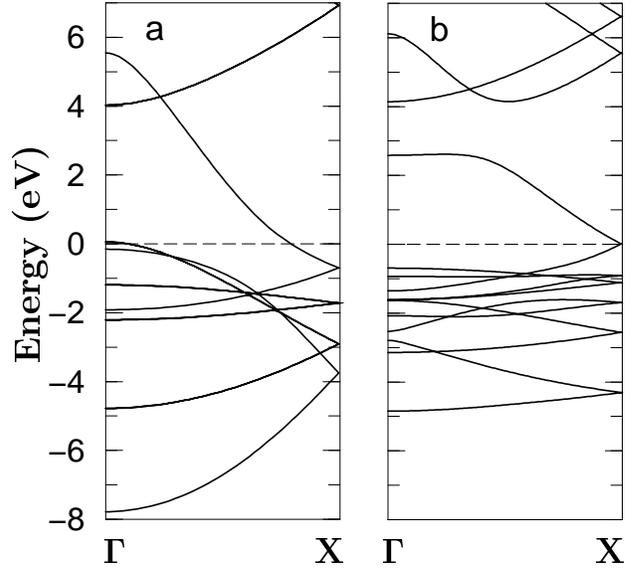}
\vspace{0.2 cm}
\caption[ ] {
   Electronic band structure of the linear (a) and
zigzag (b) wires for a length of 2.32~\AA/atom.
   The linear-wire bands have been folded onto a two-atom Brillouin
zone to facilitate the comparison.
   The energies are relative to the Fermi level.
}
\label{bands}
\end{figure}
\begin{figure}[b!]
\narrowtext
\centering
\epsfxsize=0.95\linewidth
\epsffile{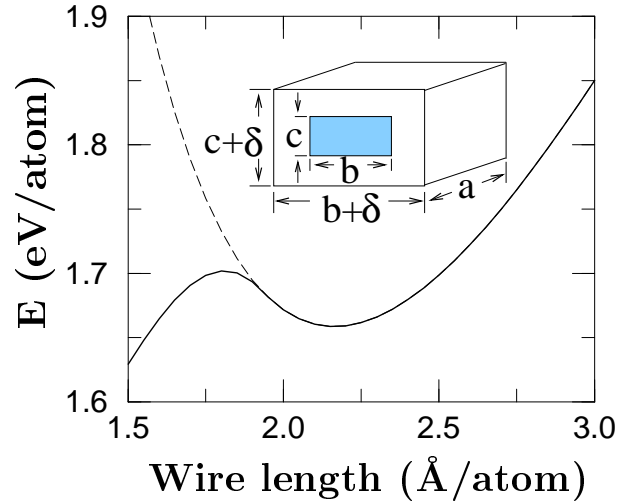}
\vspace{0.5 cm}
\caption[ ] {
   Energy versus wire length for a simple model of the wire, considered
as a free-electron tube of fixed volume, as shown in the inset,
with $a_0=\sqrt{ab}=3$~\AA\ and $\delta=2$~\AA.
   Dashed and solid lines correspond to allowing one or two
occupied bands respectively.
}
\label{modelwire}
\end{figure}

\par
   The comparison between the band structures of the linear and 
zigzag wires (Fig.~\ref{bands}) offers some hints for understanding 
their relative stability.
   In the linear chain, the overlap between the filled $d$ states
broadens the $d$ bands until they reach the Fermi level,
destabilizing the wire with their associated high density of states.
   For the same wire length, the zigzag configuration allows a larger 
bond distance, that brings back the $d$ bands below the Fermi level 
and leaves a single $s$ band crossing it.
   This is consistent with the observation of a single conduction
channel in the monatomic wires~\cite{AgraitRuitenbeek}.
   A Peierls dimerization instability is expected since the Fermi 
wave-vector is at the edge of the two-atom Brillouin zone. 
   We have observed, however,
that the magnitude of this gap-opening instability is negligible, 
only slightly noticeable just before the wire breaks, and thus
playing no substantial role in the physics described here.

\begin{figure}[!]
\narrowtext
\centering
\hspace{2 cm}\epsfxsize=0.8\linewidth
\epsffile{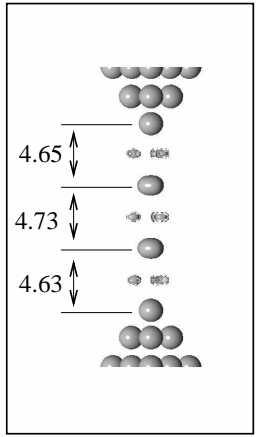}
\caption[ ] {
   Calculated valence electron density of a seven-atom wire suspended
between two small pyramidal tips, averaged over the three equivalent
relaxed configurations, with the zigzag plane rotated by 0, 120,
and 240$^\circ$ around the wire axis.
   The plot is an isosurface of constant averaged pseudo electron
density $\rho$=0.16 e/bohr$^3$.
   The numbers are relaxed distances in angstroms.
}
\label{density}
\end{figure}

\par
   Although the appearance of a zigzag instability under compression
may seem natural, its presence in a stretched wire is more surprising.
   Furthermore, its stabilization at a finite wire length is even 
harder to understand, since one would expect the wire to collapse 
into a compact, high-coordination structure typical of metals.
   However, we find that even free-standing clusters of four or 
eight atoms (the sizes calculated) are also stable with a zigzag 
chain structure.
   Although unexpected, this stability arises very naturally from the 
transverse quantization of the electron states.
   To see this, we model the wire as a tube of length $a$ per atom,
with a rectangular section $b \times c$.
   Consistently with the standard jellium 
model~\cite{Stafford-Ruitenbeek-Yannouleas-Blom}, we assume a 
fixed volume per atom $abc$, but we allow a larger `box' section
$(b+\delta) \times (c+\delta)$ to account for an electron `spillage' 
$\delta$/2 out of each jellium edge~\cite{GarciaMartin}.
   Accepting from the ab-initio calculation that the zigzag is planar, 
we also fix its thickness $c$ or, equivalently $a_0 = \sqrt{ab}$.
   The resulting free-electron energy is shown in 
Fig.~\ref{modelwire} as a function of the wire length $a$, 
for reasonable values of $a_0$ and $\delta$.
   With a single occupied band, the compromise between the transversal
and longitudinal kinetic energies results in a single minimum
(dashed line).
   Including the second band, which becomes partially occupied 
at somewhat shorter lengths, allows the energy to decrease again
(solid line), reproducing very well all the qualitative features 
observed in the ab-initio curve, such as the positions of the maximum, 
the minimum, and the point at which the second band crosses the 
Fermi level (1.83~\AA/atom).
   The basic physics that this model illustrates is the higher
stability of certain wire sections, due to the transverse
quantization of the delocalized electron 
states~\cite{Stafford-Ruitenbeek-Yannouleas-Blom}.
   This shell structure effect, which has been recently observed 
for sodium wires~\cite{Yanson99},
is similar to the so-called magic numbers (particularly 
stable sizes) of small metal clusters~\cite{Knight,deHeer}. 
   The zigzag shape is a particular realization of these stable 
sections for the monatomic gold wires.

\begin{figure}
\narrowtext
\centering
\epsfxsize=0.95\linewidth
\epsffile{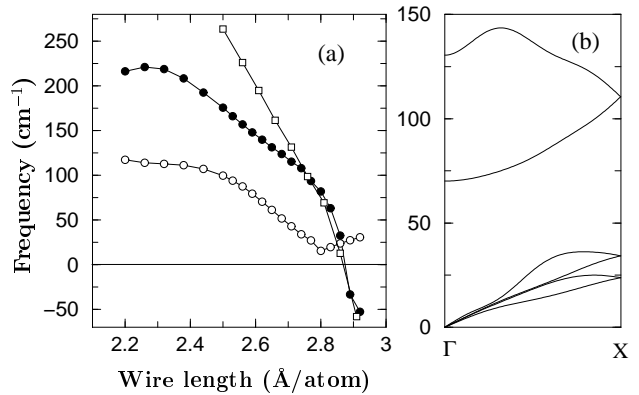}
\vspace{0.5 cm}
\caption[ ] {
   (a) Wire-length dependence of the optical phonon frequencies of
a zigzag wire at $\Gamma$, calculated using the frozen-phonon method.
   Solid and open circles stand for the longitudinal and
transversal modes respectively.
   For comparison, we also show the longitudinal phonon frequency
at X (which folds to $\Gamma$ in a two-atom cell) for a wire
constrained to be linear (squares).
   (b) Phonon dispersion curves for a zigzag wire of
2.62~\AA/atom length.
}
\label{phonons}
\end{figure}

\par
   In agreement with previous ab initio calculations~\cite{Torres99}
we find that the wire becomes unstable and breaks spontaneously 
when pulled by a force of more than 2.2 nN, i.e.\ beyond a length  
of 2.9~\AA/atom, much shorter than that apparently observed in 
stable wires~\cite{Takayanagi}.
   We offer here an explanation for this puzzling 
discrepancy, based on the predicted zigzag geometry:
if the actual wires observed have an odd number of atoms, 
with those at the extremes fixed by the contacts, the odd-numbered 
atoms would stay almost fixed on the same axis, while the even-numbered
ones could rotate rapidly around that axis, offering a fuzzy
image that could be missed by the TEM.
   We have calculated the relaxed geometry and the rotation energy 
barrier for a seven-atom wire suspended between two pyramidal tips.
   We find that the stable geometry is almost equal to that of the
infinite wire, and that the rotation barrier is only 60~meV for 
the entire wire.
   The effect is illustrated in Fig.~\ref{density}, where we 
show the electron density averaged over rotated configurations.
   Although not directly comparable to a TEM image, it can indeed
be qualitatively appreciated that the odd-numbered atoms appear
much sharper than the even-numbered ones, giving 
the impression of a four-atom wire with a large interatomic 
separation, similar to that observed experimentally.
   From the energy barrier obtained, we estimate that the 
thermal rotation would slow down to the millisecond scale,
allowing the zigzag visualization, 
only for temperatures below $\sim$40~K.

\par
   Fig.~\ref{phonons}(a) shows the calculated transversal and 
longitudinal phonon frequencies at $\Gamma$, for the zigzag wire, 
as a function of its length.
   Negative values indicate modes with imaginary 
frequency, implying the breaking of the unstable wire. 
   At the wire's equilibrium length (2.32~\AA/atom), 
the $\Gamma$-point frequencies are 113 and 219~cm$^{-1}$, 
for the transversal and longitudinal modes, respectively. 
   These are quite larger than the bulk phonon frequencies,
but comparable to those of the dimer. 
   This is not surprising if we consider that the wire interatomic 
distance is only slightly larger than that in Au$_2$. 
   Fig.~\ref{phonons}(b) shows the phonon dispersion relations
for a wire length of 2.62~\AA/atom, obtained from the
full dynamical matrix in a supercell of sixteen atoms, 
calculated with finite differences.~\cite{Ordejon95}
   We hope that the comparison of the results in 
Fig.~\ref{phonons}(a) and (b) with those of point contact 
spectroscopy experiments~\cite{Untiedt} will help to 
confirm our predicted zigzag distorsion. 

\begin{acknowledgements}
We thank N.\ Agra\"{\i}t, C.\ Balb\'as, N.\ Garc\'{\i}a, J.\ Kohanoff, 
G.\ Rubio, J.\ J.\ S\'aenz, J.\ A.\ Torres, and E.\ Tosatti 
for useful discussions.
D.\ S.\ P.\ is grateful to R. Martin for advice and support.
This work was supported by grants from Spain's DGES~PB95-0202,
and USA's DOE~8371494 and DEFG~02/96/ER~45439.
\end{acknowledgements}

\end{document}